\documentclass[sigconf]{acmart}

\AtBeginDocument{%
  \providecommand\BibTeX{{%
    \normalfont B\kern-0.5em{\scshape i\kern-0.25em b}\kern-0.8em\TeX}}}

\setcopyright{acmcopyright}
\copyrightyear{2018}
\acmYear{2018}
\acmDOI{10.1145/1122445.1122456}

\copyrightyear{2021}
\acmYear{2021}
\setcopyright{acmcopyright}\acmConference[WI-IAT '21 Companion]{IEEE/WIC/ACM International Conference on Web Intelligence}{December 14--17, 2021}{ESSENDON, VIC, Australia}
\acmBooktitle{IEEE/WIC/ACM International Conference on Web Intelligence (WI-IAT '21 Companion), December 14--17, 2021, ESSENDON, VIC, Australia}
\acmPrice{15.00}
\acmDOI{10.1145/3498851.3498934}
\acmISBN{978-1-4503-9187-0/21/12}

\begin{document}

\title{Characterizing the Anti-Vaxxers' Reply Behavior\\ on Social Media}

\author{Kunihiro Miyazaki}
\email{kunihirom@acm.com}
\affiliation{%
  \institution{The University of Tokyo}
  \city{Tokyo}
  \country{Japan}
}

\author{Takayuki Uchiba}
\email{takayuki.uchiba@sugakubunka.com}
\affiliation{%
  \institution{Sugakubunka}
  \city{Tokyo}
  \country{Japan}
}

\author{Kenji Tanaka}
\email{tanaka@tmi.t.u-tokyo.ac.jp}
\affiliation{%
  \institution{The University of Tokyo}
  \city{Tokyo}
  \country{Japan}
  }

\author{Kazutoshi Sasahara}
\email{sasahara.k.aa@m.titech.ac.jp}
\affiliation{%
  \institution{Tokyo Institute of Technology}
  \city{Tokyo}
  \country{Japan}
}


\begin{abstract}
Although the online campaigns of anti-vaccine advocates, or anti-vaxxers, severely threaten efforts for herd immunity, their reply behavior—-the form of directed messaging that can be sent beyond follow-follower relationships--remains poorly understood. 
Here, we examined the characteristics of anti-vaxxers’ reply behavior on Twitter to attempt to comprehend their characteristics of spreading their beliefs in terms of interaction frequency, content, and targets. 
Among the results, anti-vaxxers more frequently conducted reply behavior with other clusters, especially neutral accounts. 
Anti-vaxxers’ replies were significantly more toxic than those from neutral accounts and pro-vaxxers, and their toxicity, in particular, was higher with regard to the rollout of vaccines. 
Anti-vaxxers’ replies were more persuasive than the others in terms of the emotional aspect, rather than linguistical styles. 
The targets of anti-vaxxers’ replies tend to be accounts with larger numbers of followers and posts, including accounts that relate to health care or represent scientists, policy-makers, or media figures or outlets. 
We discussed how their reply behaviors are effective in spreading their beliefs, as well as possible countermeasures to restrain them. 
These findings should prove useful for pro-vaxxers and platformers to promote trusted information while reducing the effect of vaccine disinformation.
\end{abstract}

\ccsdesc[300]{Applied computing~Sociology}
\ccsdesc[300]{Computing methodologies~Machine learning}
\ccsdesc[300]{Social and professional topics}
\ccsdesc[500]{Social and professional topics~Surveillance}

\keywords{Social Media, 
COVID-19,
Vaccine,
Reply,
Toxicity}


\maketitle

\section{Introduction}

The growing anti-vaccine, or anti-vax, movement on social media critically threatens efforts to minimize the global spread of COVID-19. 
In general, suppressing pandemics requires herd immunity, which itself requires a high rate of vaccination in the population.
For example, measles requires a vaccination rate of up to 95\%~\cite{burki2019vaccine}, and COVID-19 is expected to require one of 67\%~\cite{fontanet2020covid}.
Counter to that goal, however, anti-vaxxers actively engage in propaganda activities, often online, as a means to spread their beliefs and, as a result, put herd immunity at risk~\cite{burki2019vaccine}.

Studies that have addressed the demographics of anti-vax groups and patterns in their assertions have identified types of individuals who produce anti-vaccine content~\cite{herasimenka2021understanding}.
Johnson et al.~\shortcite{johnson2020online} found that anti-vaxxers' narratives on Facebook are more attractive to users than the narratives of pro-vaxxers. 
Moreover, Germani and Biller-Andorno~\shortcite{germani2021anti}, who recently investigated the posts of anti-vaxxers on Twitter, observed that anti-vaxxers share conspiracy theories and make use of emotional language more frequently than pro-vaxxers.
Well before that, following an analysis of patterns in anti-vaccine advocacy, Kata~\shortcite{kata2012anti} proposed typologies in anti-vaxxer discourse such as ``Skewing the science'' and ``Attacking the opposition.''
Those studies and their findings are useful for identifying potential types of anti-vaxxers and the types of arguments that need to be mobilized when combating anti-vaccine beliefs.

Despite that overview, it is problematic that the reply behaviors of anti-vaxxers have not been fully investigated.
Reply behaviors, unlike normal posts or shares, essentially function as a direct means to express opinions to other users on social media.
This function is especially important when it comes to protecting people from anti-vaxxers.
This is because social media users usually follow almost only the people they want to follow and see only what they want to see~\cite{sasahara2020social}, but reply behavior can send messages beyond the follow-follower relationships~\cite{choi2020rumor}\footnote{Twitter recently added a feature that allows users to limit accounts that can reply to them, but most users still permit free reply access to them. \url{https://techcrunch.com/2020/08/11/twitter-now-lets-everyone-limit-replies-to-their-tweets/}}.
In other words, for anti-vaccine advocates, the reply behavior is a useful means to reach out to people with other beliefs.

Suppose the nature of anti-vaccine replies becomes clearer, pro-vaccine advocates will be able to prepare for anti-vaccine attacks, and platformers will be able to take measures to avoid exposing neutral people to anti-vaccine discourse.
Therefore, we analyzed the reply behavior of anti-vaxxers on Twitter and the characteristics that they adopt to increase their online presence in terms of interaction frequency, content, and targets.

Given those results, the contributions of our study are summarized as follows:
\begin{itemize}
    \item We revealed the activity patterns in the reply behavior of anti-vaxxers on Twitter;
    \item We evaluated two quantified features of anti-vaccine replies (i.e., attackness and persuasiveness);
    \item We clarified the characteristics of users and tweets that anti-vaxxers tend to target;
    \item We discussed the effectiveness of these reply behaviors with reference to previous research on propaganda and misinformation, and suggested the possible countermeasures to them.
\end{itemize}

\section{Data}
We used a longitudinal dataset of English tweets related to COVID-19 and its vaccine, lasting from February to December 2020. 
To populate the dataset, we used Twitter's Search API\footnote{\url{https://developer.twitter.com/en/docs/twitter-api/v1/tweets/search/api-reference/get-search-tweets}} to retrieve tweets containing any of the following terms related to COVID-19: ``corona virus,'' ``coronavirus,'' ``COVID19,'' ``2019-nCoV,'' ``SARS-CoV-2,'' and ``wuhanpneumonia.'' From the results, tweets containing the words ``vaccine,'' ``vax,'' or ``vaccination'' were retained for analysis. 

The resulting volume of tweets was 8,579,728, of which 6,879,713 (80.2\%) were retweets (RT) and 293,946 (3.43\%) were replies (RP). The number of unique users was 2,799,034.
We found that the ratio of replies among all tweets was considerably small. 
Reply behavior, however, is a directed messaging that can be sent regardless of follow-follower relationships.
Thus, anti-vaccine replies, even if they are few in number, can be used to socially influence recipients more frequently compared to mere exposure of undirected tweets and retweets. 
Furthermore, a certain ratio of the anti-vaccine replies was sent to users in other groups, confirming that reply behavior is used to reach out to people with other beliefs, which we will show later.

\section{Classification of anti-vax and other groups}
To identify anti-vaccine groups, we employed the RT network clustering to classify users according to their stance on vaccines. 
The RT network clustering is a method of detecting users with similar stances by applying network clustering to a retweet (RT) network~\cite{fortunato2010community,conover2011political}. 
Previous research has shown that a network community can be easily divided by stances, especially on a topic related to strong beliefs such as vaccinations~\cite{gunaratne2019temporal,cossard2020falling}. 


A RT network was created with users as nodes, and an edge consists of users with more than two RTs (including mutual RTs). 
By using more than two RTs, the meaning of the endorsement was more robustly incorporated into the edges~\cite{garimella2018quantifying}.
We used all data from February to December 2020.
We did not include quote retweets (QTs) in the RTs here because QTs often do not indicate the endorsement.
After creating a RT network, we applied $k$-core decomposition ($k=3$) to exclude users with only weak connections to the primary discussions~\cite{alvarez2006large}. 
Then, the Louvain method was used to cluster anti-vaccine users and other groups~\cite{blondel2008fast}. 
Resolution of clustering was set to 1.

\begin{figure}[htbp]
\centering
\includegraphics[width=0.95\linewidth]{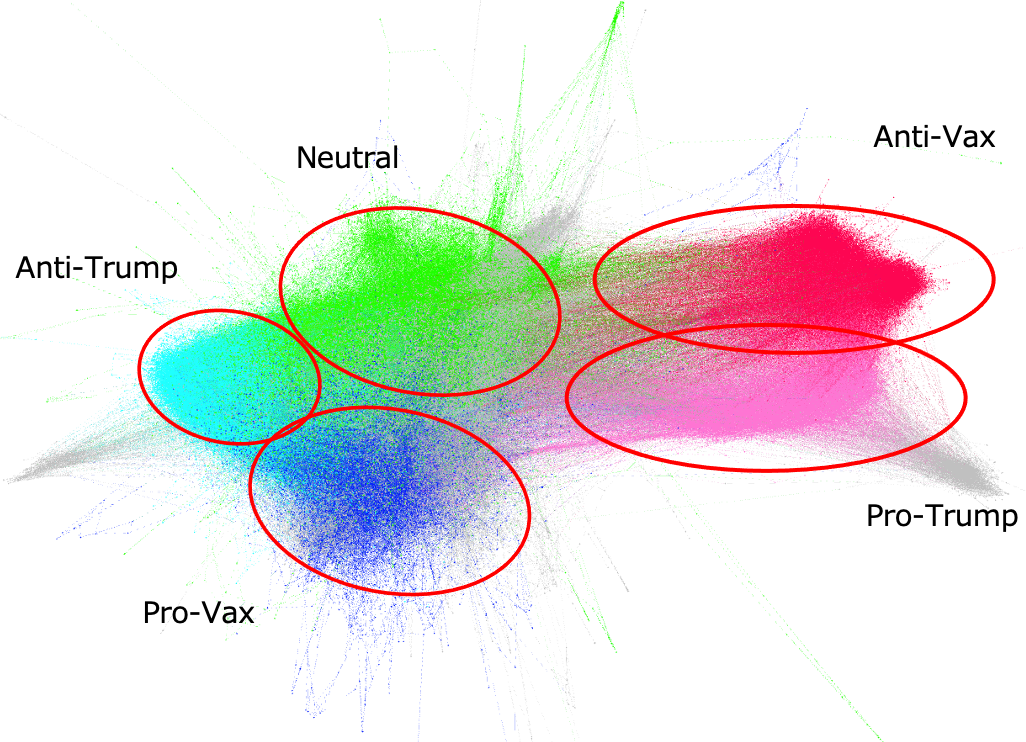}
\caption{Results of RT network clustering.}
\label{Network}
\end{figure}

Figure~\ref{Network} illustrates the results of the RT network clustering. 
The visualization tool Gephi~\cite{bastian2009gephi} was used with the graph layout algorithm ForceAtlas2~\cite{jacomy2014forceatlas2}. The number of nodes was 47,135, while the number of edges was 241,370.
Due to the constraints RT$>=$2 and $k$-core$=$3, the number of users was reduced to only those actively participating in the discussion about vaccines. 
As a result of Louvain clustering, the five largest clusters were obtained, all marked with circles in Figure~\ref{Network}. 
We labeled these clusters by scrutinizing (1) the ten most retweeted accounts, (2) the ten most retweeted tweets, and (3) the word cloud of the 100 most retweeted tweets for each cluster ((1)-(3) are not shown in the paper because of the limitation of the space).
By percentage of cluster size, the clusters were Pro-Vax (9.17\%), Anti-Trump (16.31\%), Neutral (24.84\%), Anti-Vax (12.23\%), and Pro-Trump (12.18\%). That is, we observed three clusters with a clear stance on vaccines—in favor, against, and neutral—and two clusters with strong political ideologies, both of which had a relatively clear position on vaccines (Pro-Trump is close to Anti-Vax and Anti-Trump is close to Pro-Vax). 
These five clusters were robustly obtained when changing parameters for clustering (e.g., resolution of clustering). 
Although the purpose of this study is to examine the communications of vaccine-related clusters, we included the political clusters (Pro-Trump and Anti-Trump) for analysis because we thought (1) the portion of political clusters was too big to remove and (2) politics have been involved in a large portion of the vaccine topic and were, thus, important.

\section{Interaction frequency of replies}

\begin{table*}[htbp]
\centering

\begin{tabular}{lccccccc}
\hline
Cluster    & Users & TW      & TW/Users & RP     & RP/TW(\%) & RT      & RT/TW(\%) \\
\hline
Pro-Vax   & 4,320    & 216,130 & 50.03   & 6,501  & 3.01      & 165,599 & 76.6      \\
Anti-Trump & 7,689    & 351,436 & 45.71   & 3,302  & 0.94      & 319,044 & 90.8      \\
Anti-Vax   & 5,766    & 282,665 & 49.02   & 14,291 & \textbf{5.06}      & 240,648 & 85.1      \\
Pro-Trump  & 5,739    & 203,288 & 35.42   & 1,419  & 0.70      & 188,925 & 92.9      \\
Neutral    & 11,713   & 544,882 & 46.52   & 6,800  & 1.25      & 438,701 & 80.5     \\
\hline
\end{tabular}
\caption{Statistics of the activity of each cluster. TW: tweets. RP: replies. RT: retweets. TW includes normal tweets, RP, and RT. Reply rate (RP/TW) was significantly higher in the Anti-Vax cluster (chi-square test: $p < 0.001$).}
\label{stats}
\end{table*}

We analyzed whether the frequency of reply behavior differed from cluster to cluster. 
Table~\ref{stats}, listing statistics regarding each cluster's behavior, shows that the Anti-Vax cluster had the highest ratio of replies to the number of tweets (5.06\%, shown in bold), with which a chi-square test with the sum of the other four clusters revealed was significant ($p < 0.001$)\footnote{We conducted the chi-square test to the numbers of posts, not the ratios}. 
The result suggests that the Anti-Vax cluster was more active in reply behavior. 

\begin{table}[!t]
\centering

\begin{tabular}{lccccc}
\hline
           & \multicolumn{5}{c}{Target}                             \\ \cline{2-6} 
Source     & PV       & AT       & AV       & PT       & N \\
\hline
Pro-Vax(PV)   & 86.34    & 2.93       & 4.90     & 1.45      & 4.38    \\
Anti-Trump(AT) & 4.10     & 78.87      & 0.97     & 8.00      & 8.05    \\
Anti-Vax(AV)   & 4.73     & 2.59       & \textbf{65.93}    & 8.01      & \underline{18.74}   \\
Pro-Trump(PT)  & 1.79     & 2.44       & 7.31     & 82.31     & 6.15    \\
Neutral(N)    & 1.92     & 3.91       & 1.20     & 1.85      & 91.12 
\\ \hline 
\end{tabular}
\caption{Ratio of targets of replies by cluster (unit: \%). The Anti-Vax cluster had the smallest ratio of inner-cluster replies (chi-square test: $p < 0.001$).
The main target of the Anti-Vax replies was Neutral.
}
\label{sorce_target}
\end{table}

Table~\ref{sorce_target} shows the ratio of reply behavior in terms of the targets of replies. 
Therein, every cluster has a large portion of replies from inner clusters (i.e., diagonal components in the table). 
Among them, the Anti-Vax cluster had, by far, the lowest ratio of inner-cluster replies (65.93\%, $p < 0.001$, shown in bold), meaning that it had the most frequent inter-cluster replies\footnote{Here, inter-cluster replies mean the replies to other clusters. Inner-cluster replies mean the replies to the same clusters}: approximately 34\% of all its replies. 
The main target of Anti-Vax replies was in the Neutral cluster (18.74\%, shown with underlining). 
That finding is consistent with Johnson et al.~\shortcite{johnson2020online}, whose Anti-Vax cluster interacted more with the Neutral cluster as well. 
On the other hand, the number of replies from the Anti-Vax to Pro-Vax clusters was relatively small.
Thus, the key takeaways thus far are that:

\begin{itemize}
    \item Anti-vaxxers have a higher reply rate than other users;
    \item Anti-vaxxers often reply to users outside their group; and
    \item Anti-vaxxers reply to neutral users more frequently.
\end{itemize}

\section{Contents of replies}
Previous research has revealed two characteristics of replies on social media in terms of the propagation of beliefs: attackness and persuasiveness~\cite{rossini2018social}. Whereas attackness may involve publicly vilifying political candidates in competing campaigns on social media during election season to achieve chilling effects~\cite{hua2020characterizing,hua2020towards}, persuasiveness may involve seeking to expand one's preferred campaign by altering others' beliefs~\cite{an2019political,wasike2017persuasion}. 
We, therefore, quantified the attackness and persuasiveness of the texts from Twitter to characterize the reply behaviors of anti-vaxxers.

\subsection{Methods}
As for attackness, we employed Google's Perspective API\footnote{\label{perspective}\url{https://www.perspectiveapi.com/}}, a widely used tool in recent years~\cite{hua2020characterizing,hua2020towards,wu2021cross} that allows users to measure a text's \textit{toxicity} on a scale from 0 to 1.

As for persuasiveness, following the literature~\cite{tan2016winning,an2019political}, we measured the number of words (\#words), first-person singular pronouns (1SG), first-person plural pronouns (1PL), second-person pronouns (2), number of positive words (pos.), number of negative words (neg.), number of question marks (?), level of arousal (arousal), and valence level (valence). 
For all metrics except \#words, we used the dictionaries from LIWC 2015~\cite{pennebaker2015development} and~\citeauthor{warriner2013norms}~\citeyear{warriner2013norms} and counted the number of words registered in the dictionaries that appeared in each tweet\footnote{As for the dictionary of~\cite{warriner2013norms} we counted the number of words per tweet with scores above the median for each metric in the dictionary.}. 
In addition, we added the presence or absence of URLs in replies (URL) as an indicator of evidence-based opinions~\cite{schwarz2011augmenting}.

\subsection{Results}

\begin{figure}[!t]
\centering
\includegraphics[width=0.95\linewidth]{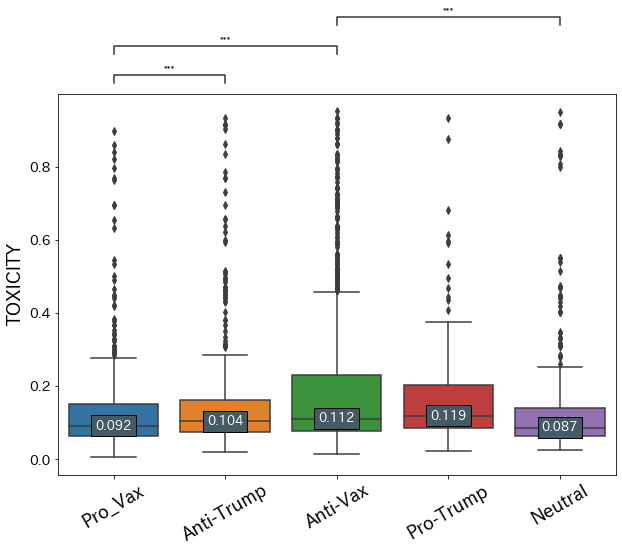}
\caption{
Toxicity scores of the inter-cluster replies for each cluster. 
Each median score is annotated in boxes.
Replies from the Anti-Vax cluster was significantly more toxic than ones from the Pro-Vax and Neutral clusters (*** $p < 0.001$ by Mann–Whitney U test with Bonferroni correction).
}
\label{boxplot}
\end{figure}

\textbf{Toxicity:}
Figure~\ref{boxplot} shows the toxicity of each inter-cluster reply of each cluster.
We found that the absolute values of toxicity in replies were relatively low; the median scores range from 0.087 to 0.112, which were far below the threshold of ``highly-adversarial'' tweets, which was once set as 0.7 in literature~\cite{hua2020characterizing,hua2020towards}.
When looking at the ratio of replies with a toxicity score above 0.7, we obtained less than 3\% for all clusters. 
In relative comparison, however, we can see that the toxicity scores of the Anti-Vax were significantly higher than the Pro-Vax and Neutral clusters, which was somewhat expected. 


From these results, we can see that: (1) the inter-cluster replies of Anti-Vax were overall more toxic than those from Pro-Vax and Neutral users, although the ratio of highly toxic replies is under 3\%; 
(2) therefore, when communicating with anti-vaxxers, one must be careful with the anti-vaxxers’ more-toxic-than-others replies while keeping in mind that extremely strong attacks do not often occur.
Especially, considering that Anti-Vax users were replying most frequently to the Neutral cluster, the replies that tend to be toxic might have a non-negligible effect on them.

In addition, to gauge the attackness to each cluster, we further divided the inter-cluster replies by their targeted clusters (Table~\ref{toxicity_s_t}). 
According to the Mann–Whitney U test with Bonferroni correction, among inter-cluster replies from the Anti-Vax cluster, only the attackness to the Pro-Trump cluster was significantly higher than those to the other clusters (0.15, $p < 0.001$, shown in bold). 
This result was unexpected because we initially hypothesized that the Pro-Vax cluster would be the primary target of the Anti-Vax cluster's attacks, with the goal of achieving a chilling effect.
Upon scrutinization of the content of the highly toxic replies from Anti-Vax to Pro-Trump, we found most of the replies to be among the lines of ``Don't spread the vaccine.'' 
As the Trump administration was in power, it seems that complaints and requests were made against the administration with high attackness, even though the Pro-Trump cluster was close to the Anti-Vax cluster in the RT network.
From this, we can see that the toxic replies from Anti-Vax were mainly towards the administration operating the vaccination nationwide rather than the pro-vaxxers, who are advocates of vaccination.

\begin{table}[htbp]
\centering

\begin{tabular}{lccccc}
\hline
                & \multicolumn{5}{c}{Target}   \\ \cline{2-6} 
Source & PV    & AT    & AV    & PT    & N     \\ \hline
Pro-Vax(PV)     & -     & 0.09 & 0.10 & 0.08 & 0.09 \\
Anti-Trump(AT)  & 0.09 & -     & 0.10 & 0.11 & 0.11 \\
Anti-Vax(AV)    & 0.11 & 0.10 & -     & \textbf{0.15} & 0.11 \\
Pro-Trump(PT)   & 0.16 & 0.13 & 0.11 & -     & 0.13 \\
Neutral(N)      & 0.08 & 0.09 & 0.09 & 0.09 & -     \\ \hline
\end{tabular}
\caption{
Median toxicity scores for inter-cluster replies for each cluster by targets. The scores for inner-cluster replies are masked. Replies from the Anti-Vax cluster were especially toxic for the Pro-Trump cluster (shown in bold). 
}
\label{toxicity_s_t}
\end{table}

\textbf{Persuasiveness:}
Table~\ref{persuasion} summarizes the results of the Mann–Whitney U test with Bonferroni correction for the metrics of persuasiveness described in the Methods.
For the visibility, we only show the metrics in which Anti-Vax showed a significant difference in the direction of high persuasiveness. 
From this table, we can see that the metrics only related to emotions showed the high persuasiveness for Anti-Vax, i.e., low positivity, high negativity, low valence, and low arousal.
As for the other metrics, we could not see a significantly high persuasiveness for Anti-Vax; rather, Anti-Vax was less persuasive in some metrics, e.g., 1PL.
From this result, we found that the replies from anti-vaxxers are more persuasive in terms of emotional expressions rather than linguistical styles.

\begin{table*}[htbp]
\centering

\begin{tabular}{lcccccccccc}
\hline
\multicolumn{1}{c}{Cluster pairs}               & pos.                              & neg.                                       & valence                           & arousal                           \\ \hline
Anti-Vax vs Pro\_Vax   & \textless{}\textless{}\textless{} & \textgreater{}\textgreater{}\textgreater{} & \textless{}                       & -                                 \\
Anti-Vax vs Neutral    & \textless{}\textless{}\textless{} & \textgreater{}\textgreater{}\textgreater{} & \textless{}\textless{}\textless{} & -                                 \\
Anti-Vax vs Pro-Trump  & -                                 & \textgreater{}\textgreater{}               & -                                 & -                                 \\
Anti-Vax vs Anti-Trump & \textless{}\textless{}\textless{} & -                                          & \textless{}\textless{}\textless{} & \textless{}\textless{}\textless{}
 \\ \hline

\end{tabular}
\caption{Comparison of persuasiveness between each cluster. The comparison is based on each feature using the Mann-Whitney U test with Bonferroni correction. 
The direction of brackets indicates the cluster with the larger amount, e.g., the direction of ``$>$'' shows that Anti-Vax has a higher score for the item in question.
``$<<<$'': $p < 0.001$, ``$<<$'': $p < 0.01$, ``$<$'': $p < 0.05$.
}
\label{persuasion}
\end{table*}

In sum, the results suggest that:
\begin{itemize}
    \item The inter-cluster replies of anti-vaxxers are significantly more toxic than those of pro-vaxxers and neutral users, although the extremely high toxic replies are rare;
    \item The toxicity of anti-vaxxers seems to be mainly towards the administration in charge of rolling out the vaccines rather than pro-vax advocates;
    \item Regarding persuasiveness, anti-vaxxers use more emotional expressions, including negative words, than pro-vaxxers and neutral users;
    \item Linguistic styles (e.g., word length and use of the first person) do not significantly differ between anti-vaxxers and other clusters; and
\end{itemize}

\section{Target of anti-vaxxers' reply}
One of the unique features of reply behavior is that its message always has a target.
Here, we analyzed the characteristics of users and tweets that are likely to receive replies from anti-vaxxers, expecting to identify who should be cautious around anti-vaxxers' replies.
For the analysis, we used basic user and tweet information obtained via Twitter's API.

\textbf{User features:}
For the user features, we used Twitter statistics and information from user bios. Twitter statistics included (1) the number of followers, an indicator of the account’s presence; (2) account age (in days), indicating whether the ac- count was recently created; and (3) status count per day, meaning how often the account posts tweets. Concerning information from the bios, we used (4) bio length, how appealing an account appears to be, and (5) the usage of words in the bio for the qualitative analysis of accounts.

The results appear in Table~\ref{twitter_stats}. 
The table shows that receivers tended to have significantly larger values in all metrics. 
That is, users who received replies had more followers, were older accounts, were usually more active in posting tweets, and were more appealing accounts than non-receivers. 
In particular, their median number of followers was 19 times greater than other accounts. 
It seems reasonable that accounts that are large and post many times stand out, making them easy targets for anti-vaccine groups.

\begin{table}[htbp]

\centering
\begin{tabular}{lcccc}
\hline
             & Receivers & Non-receivers & test \\
             \hline
followers    & 194,652   & 1,141         & \textgreater{}\textgreater{}\textgreater{}  \\
statuses/age & 16.34 & 12.73     & \textgreater{}\textgreater{}\textgreater{} \\
age          & 4,081     & 2,392         & \textgreater{}\textgreater{}\textgreater{}  \\
bio length   & 18       & 15           & \textgreater{}\textgreater{}\textgreater{}  \\
\hline
\end{tabular}
\caption{
Twitter statistics (medians) of accounts that received replies from users in the Anti-Vax cluster (Receivers) and those that did not (Non-receivers). ``Test'' indicates the results of the Mann–Whitney U test showing that Receivers had larger values for all statistics, with $p < 0.001$ (``$>>>$'').}
\label{twitter_stats}
\end{table}

Regarding the difference in the content of bios, 38 words were significantly more common among receivers (chi-square test: p $<$ 0.05 for all\footnote{For the chi-square test, we created a 2 x 2 matrix for each word composed of (replied and non-replied tweets) x (tweets with and without the word).}). We categorized those words into nine categories: 
(1) Accounts related to health care: ``health'' and ``care;''
(2) Scientists: ``professor;''
(3) Policymakers: ``house'' and ``governor;''
(4) Writers: ``journalist,'' ``editor,'' ``correspondent,'' ``author,'' and ``reporter;''
(5) News media: ``CNN,'' ``news,'' ``story,'' ``newsletter,'' ``subscribe,'' ``politics,'' ``tip,'' ``update,'' ``business,'' ``analysis,'' ``sport,'' ``team,'' and ``source;''
(6) Media personalities: ``host'' and ``podcast;''
(7) Self-branding: ``dad,'' ``husband,'' ``view,'' ``opinion,'' and ``endorsement;''
(8) Representatives of organizations: ``founder,'' ``president,'' ``director,'' and ``member;''
(9) Official account of organizations: ``Twitter,'' ``account,'' ``policy,'' and ``tweet.''
Media-related accounts often received replies from the Anti-Vax cluster. It is understandable that accounts that relate to health care or represent scientists or policymakers were common targets of the cluster.

\textbf{Tweet features:}
Next, we examined whether the characteristics of the tweets targeted by the Anti-Vax cluster differed from those that were not. 
For the quantitative analysis, we used 
(1) the indicators of attackness and persuasiveness already used in the previous analysis; 
(2) the number of RTs, which indicates conspicuousness;
(3) whether the tweet was a reply, which indicates if the tweet was contextualized in a discussion; and 
(4) the number of words in the tweets for qualitative analysis.

Table~\ref{targeted_tweet} shows the quantitative analysis results.
We can see significant differences in almost all the items. 
It is interesting that highly toxic tweets receive fewer replies from anti-vaxxers, although it is difficult to obtain any practical implications from this finding. 
On the other hand, tweets with more words, richness in emotions (pos., neg., valence, and arousal), and high persuasiveness in linguistics (e.g., 1SG) tend to receive more replies from anti-vaxxers.


When looking at the bottom part of Table~\ref{targeted_tweet}, tweets with more retweets were more likely to receive replies from anti-vaxxers, which seems to be reasonable considering that tweets with more replies simply stand out more. In addition, we could not see a difference between replies and retweets.

\begin{table}[htb]

\centering
\begin{tabular}{lc}
\hline
         & Targets vs. non-targets                    \\ \hline
Toxicity & \textless{}\textless{}\textless{}          \\
\#words  & \textgreater{}\textgreater{}\textgreater{} \\
pos.     & \textgreater{}\textgreater{}\textgreater{} \\
neg.     & \textgreater{}\textgreater{}\textgreater{} \\
1SG      & \textgreater{}\textgreater{}\textgreater{} \\
1PL      & \textgreater{}\textgreater{}\textgreater{} \\
2        & \textgreater{}\textgreater{}\textgreater{} \\
?        & \textgreater{}                             \\
valence  & \textgreater{}\textgreater{}               \\
arousal  & \textgreater{}\textgreater{}\textgreater{} \\
URL      & \textgreater{}\textgreater{}\textgreater{} \\
\hline
retweet & \textgreater{}\textgreater{}\textgreater{} \\
reply    & -                                          \\ \hline
\end{tabular}
\caption{
The difference of targeted tweets (targets) and non-targeted tweets (non-targets). The difference was tested in terms of each item in the left column. The direction of the bracket indicates the larger amount of the item (e.g., toxicity is higher for non-target). The number of brackets shows the significance of difference; ``$>>>$'': $p < 0.001$, ``$>>$'': $p < 0.01$, ``$>$'': $p < 0.05$. The chi-square test was used for URL retweets and replies, while the Mann–Whitney U test was used for all other items.
}
\label{targeted_tweet}
\end{table}


Last, we analyzed common words in targeted tweets. We used the same method as in the analysis of the content of users' bios. As a result, we acquired 16 words, categorized in the following themes:
(1) Cases of COVID-19: ``death,'' ``life,'' ``case,'' and ``people;''
(2) Policymakers: ``Fauci,'' ``president,'' and ``government;''
(3) Operation of government: ``vaccination,'' ``lockdown,'' ``test,''; ``way,'' ``mask,'' ``immunity,'' and ``jab;''
(4) Child: ``child;''
(5) Call to action: ``everyone;''
(6) Comparison with flu: ``flu.''
Except for the case of COVID-19, the topics that attracted the replies of anti-vaxxers were about the government and its operations, children, and a call to action for all.
Criticism of the government is reasonable considering the result of reply frequency in this study that Pro-Trump receives more toxic replies from Anti-Vax.
As for children, previous studies have also reported a high level of anti-vaxxers' interest in children matters~\cite{kata2012anti}.

Altogether, the analysis of reply targets suggests that:
\begin{itemize}
    \item Users most likely to receive replies from the Anti-Vax cluster have large numbers of followers, have consistently posted messages for a long time, and have longer bios;
    \item Accounts most likely to receive replies are related to health care, academia, policymaking, and media; and
    \item In terms of content, tweets addressing cases of COVID-19, policymakers and their operations, and children are more likely to get replies from anti-vaxxers.
\end{itemize}

\section{Discussion}
Here, we discuss how the reply behaviors of anti-vaxxers are effective in terms of spreading beliefs and possible countermeasures to those behaviors.

\subsection{Effectiveness of replies}
\textbf{Highly frequent replies:}
Repeated exposure to the same belief is one of the most effective ways to affect people’s perception in what is known as the ``mere exposure effect'', coupled with subliminal techniques often used in propaganda~\cite{bornstein1989subliminal,bornstein1992stimulus}.
From that viewpoint, the efforts of anti-vaxxers to frequently send messages to other clusters are a threat to social media users.

\noindent
\textbf{Highly emotional replies:}
Previous research on effective vaccine narratives has shown that texts with strong emotion were more likely to leave a greater impression on receivers than texts high in richness or detailed expression~\cite{betsch2011influence}.
Other researchers have reported that influential users on social media tend to be individuals who express negative sentiments~\cite{quercia2011mood,xiao2019changing}.
Similarly, our study showed that the replies of anti-vaxxers conveyed more negative emotions (i.e., with higher toxicity), although their replies were not linguistically persuasive. 
That finding suggests that the style of anti-vaxxers’ messages aligns with effective means of propaganda.

\noindent
\textbf{Replies to prominent accounts:}
Borrowing the authority of prominent social media accounts is a typical way of spreading disinformation, as demonstrated by Russian manipulations during the 2016 U.S. presidential election~\cite{benkler2018network}.
When it comes to spreading anti-vaccine beliefs, replying to prominent social media accounts can generate significant impressions from other accounts because Twitter’s current settings allow users to easily see replies as they jump to the page of a particular tweet, which is considered to be an effective way to promote their beliefs to others.

\subsection{Implications for combating anti-vax propaganda.}

First, pro-vaxxers might have to reach out to other groups more than before to prevent the effort of anti-vaxxers from bearing fruit. 
This argument has also been made in the literature; for example,~\citeauthor{johnson2020online}~\shortcite{johnson2020online} showed that the presence of pro-vaxxers is less to neutral people than one of anti-vaxxers, and~\citeauthor{burki2019vaccine}~\shortcite{burki2019vaccine} cited that the medical community needs to be more proactive. 
Although there is a risk that simply responding to anti-vaccine arguments can be counterproductive, using elaborate ways to fight disinformation might be effective, such as actively introducing official information to people searching for it~\cite{burki2019vaccine}.

Second, social media users should prepare for emotional replies from anti-vaxxers.
In the interviews that \citeauthor{steffens2019organisations} ~\shortcite{steffens2019organisations} conducted with pro-vax organizations about their experiences with responding to anti-vaxxers, an interviewee highlighted the ``need to come across as the responsible, reasonable, calm ones because of all the people that are reading and not commenting.'' 
\citeauthor{steffens2019organisations} also found that anti-vaxxers were relatively negative in their expressions and tone. Those findings, including ours, should be shared by social media users.

Third, the platformer may need to lower the visibility of the replies from anti-vaxxers to prominent accounts.
This can be mitigated by lowering the order of the anti-vaxxers' replies in the replies displayed to users.

\subsection{Limitations and Future Work}

\noindent
\textbf{Limitations of the Perspective API:}
As earlier work has suggested~\cite{hua2020characterizing}, the Perspective API does not include Twitter data in its training data, which may cause errors in prediction. 
However, we believe that the limitation should not affect our statistical comparisons, because we used the API with a sufficient number of tweets from clusters rather than focused on a single specific tweet.

\noindent
\textbf{Bot analysis:}
Previous studies have revealed that bots (i.e., automated accounts) are active in discourse about vaccines on social media platforms~\cite{broniatowski2018weaponized}.
Future work should involve analyzing the reply behavior of bot accounts and their impact on vaccine-related propaganda.

\noindent
\textbf{Differences from Reddit:}
Regarding persuasiveness, previous studies have used Reddit as the examined platform, whereas we used Twitter.
A major difference between them is that Reddit allows longer messages, whereas Twitter allows only phrases with 280 characters or under in English. 
The platform’s difference may affect persuasiveness, especially in terms of linguistic style. 
For example, we introduced the number of words as a measure for comparison, which might have made it difficult to discern any significant difference on Twitter. 
In addition, other measures (e.g., use of the first person) may have been reduced relative to Reddit in order to convey essential information only. 

\section{Conclusion}
We analyzed the reply behavior (i.e., the directed messaging) of anti-vaxxers on Twitter in search of insights into their spreading beliefs as well as counterstrategies to restrain it. 
According to our results, the anti-vax's reply behavior is characterized by tweets involving strong emotion with toxic words and/or persuasion with negative words.
In particular, their toxicity was higher when it comes to the rollout of vaccines. Anti-vaxxers were shown to make frequent replies, often targeting prominent accounts with large numbers of followers. These results suggest some policies for a counterstrategy to anti-vaxxers. 
Prominent accounts in pro-vaccine and neutral groups with more followers and constant postings should be prioritized to receive guidance for countering the replies of anti-vaxxers. 
Such counterstrategies to design social media platforms and conduct fact-checking are essential to overcoming the COVID-19 pandemic and infodemic.

\begin{acks}
This work was supported by JST, CREST Grant Number JPMJCR20D3, Japan.
\end{acks}
\bibliographystyle{ACM-Reference-Format}
\bibliography{sample-base}

\end{document}